\documentclass[a4paper,12pt]{article}
\usepackage{graphicx, color}
\usepackage{latexsym}
\begin{document}
\renewcommand{\baselinestretch}{1.}
\def\BA{\Bbb A}
\def\BC{\Bbb C}
\def\BR{\Bbb R}
\newfont{\titlefont}{cmssbx10 scaled\magstep5}
\newcommand{\CA}{{\cal A}}
\newcommand{\CB}{{\cal B}}
\newcommand{\CC}{{\cal C}}
\newcommand{\CD}{{\cal D}}
\newcommand{\CE}{{\cal E}}
\newcommand{\CF}{{\cal F}}
\newcommand{\CG}{{\cal G}}
\newcommand{\CH}{{\cal H}}
\newcommand{\CI}{{\cal I}}
\newcommand{\CJ}{{\cal J}}
\newcommand{\CK}{{\cal K}}
\newcommand{\CL}{{\cal L}}
\newcommand{\CM}{{\cal M}}
\newcommand{\CN}{{\cal N}}
\newcommand{\CO}{{\cal O}}
\newcommand{\CP}{{\cal P}}
\newcommand{\CQ}{{\cal Q}}
\newcommand{\CR}{{\cal R}}
\newcommand{\CS}{{\cal S}}
\newcommand{\CT}{{\cal T}}
\newcommand{\CU}{{\cal U}}
\newcommand{\CV}{{\cal V}}
\newcommand{\CW}{{\cal W}}
\newcommand{\CX}{{\cal X}}
\newcommand{\CY}{{\cal Y}}
\newcommand{\CZ}{{\cal Z}}
\newcommand{\bea}{\begin{eqnarray}} \newcommand{\eea}{\end{eqnarray}}
\newcommand{\beqa}{\begin{eqnarray}} \newcommand{\eeqa}{\end{eqnarray}}
\newcommand{\beq}{\begin{equation}} \newcommand{\eeq}{\end{equation}}
\newcommand{\non}{\nonumber} \newcommand{\eqn}[1]{\beq {#1}\eeq}
\newcommand{\eu}{Euclidean\ }
\newcommand{\lmk}{\left(} \newcommand{\rmk}{\right)}
\newcommand{\lkk}{\left[}
\newcommand{\rkk}{\right]} \newcommand{\lhk}{\left \{ }
\newcommand{\rhk}{\right \} } \newcommand{\lnk}{\left \{ }
\newcommand{\rnk}{\right \} } \newcommand{\del}{\partial}
\newcommand{\abs}[1]{\vert{#1}\vert}
\newcommand{\vect}[1]{\mbox{\boldmath${#1}$}} \newcommand{\bib}{\bibitem}
\newcommand{\new}{\newblock}
\newcommand{\la}{\left\langle} \newcommand{\ra}{\right\rangle}
\newcommand{\bfx}{{\bf x}} \newcommand{\bfk}{{\bf k}}
\newcommand{\vex}{{\vect x}}
\newcommand{\vek}{{\vect k}} \newcommand{\vep}{{\vect p}}
\newcommand{\veq}{{\vect q}} \newcommand{\vev}{{\vect v}}
\newcommand{\vej}{{\vect j}} \newcommand{\veg}{{\vect \gamma}}
\newcommand{\vena}{{\vect \nabra}} \newcommand{\vebt}{{\vect \beta}}
\newcommand{\gtilde}{~\mbox{\raisebox{-1.0ex}{$\stackrel{\textstyle >}
{\textstyle \sim}$ }}}
\newcommand{\ltilde}{~\mbox{\raisebox{-1.0ex}{$\stackrel{\textstyle <}
{\textstyle \sim}$ }}}
\newcommand{\gsim}{~\mbox{\raisebox{-1.0ex}{$\stackrel{\textstyle >}
{\textstyle \sim}$ }}}
\newcommand{\lsim}{~\mbox{\raisebox{-1.0ex}{$\stackrel{\textstyle <}
{\textstyle \sim}$ }}}
\newcommand{\mh}{m_H} \newcommand{\veff}{V_{\rm eff}}
\newcommand{\phip}{\phi_+}
\newcommand{\eff}{\rm eff}
\newcommand{\Ghat}{\hat{\Gamma}}
\newcommand{\mpl}{M_{Pl}}
\newcommand{\gn}{g_{*N}}
\title{A Calculation of Baryon Diffusion Constant \\
in Hot and Dense Hadronic Matter \\
Based on an Event Generator URASiMA}

\author{N.\ Sasaki, O.\ Miyamura, 
S.\ Muroya\footnote{E-MAIL:muroya@yukawa.kyoto-u.ac.jp , TEL: (0834)-29-2007, FAX:(0834)-29-2058}
and C.\ Nonaka \\ 
{\it Department of Physics, Hiroshima University, }\\
{\it Higashi-Hiroshima 739-8526, Japan }\\
{\it $^*$Tokuyama Women's College, Tokuyama, 745-8511, Japan } }

\maketitle

\begin{abstract}
We evaluate thermodynamical quantities and transport coefficients of a dense
and hot hadronic matter based on an event generator URASiMA
(Ultra-Relativistic AA collision Simulator based on Multiple
Scattering Algorithm).
The statistical ensembles in equilibrium with fixed temperature and chemical
potential are generated by imposing periodic boundary condition to the
simulation of URASiMA, where energy density and baryon number density is
conserved.
Achievement of the thermal equilibrium and the chemical equilibrium are
confirmed by the common value of slope parameter in the energy
distributions
and the
saturation of the numbers of contained particles, respectively.
By using the generated ensembles, we investigate the temperature dependence
and the chemical potential dependence of the baryon diffusion constant of a
dense and hot hadronic matter.
\end{abstract}

\vspace{1.5cm}

\newpage
\baselineskip 24pt
\section{ Introduction} \label{intr}

Physics of a high density and high temperature hadronic matter has been
highly
attracting in the context of both high energy nuclear collisions and
cosmology
as well as theoretical interest\cite{QM99}. In the recent ultra-relativistic
nuclear collisions, though the main purpose should be confirmation of
Quark-Gluon Plasma(QGP) state, physics of hot and/or dense hadronic state
dominates the system. Hence, thermodynamical properties and transport
coefficients of a hadronic matter are essentially important for the
phenomenological description of the space-time evolution of the produced
exited region. In the cosmology, in addition to the global evolution of the
early universe, baryon diffusion would play an important roll in the
nucleosynthesis problem.

Because of the highly non-perturbative property of a hot and dense hadronic
state, investigation on the thermodynamical properties and transport
coefficients has been hardly investigated.  Numerical simulation
based on Lattice gauge theory is a very powerful tool for the analysis of
finite temperature QCD. Recently, transport coefficient of hot gluonic
matter
has been investigated \cite{Sakai}.
But even for the modern high-performance super-computer, lattice QCD
evaluation of the transport coefficients of hadronic matter is very
difficult, especially below $T_c$. Furthermore, at finite density, present
numerical scheme of lattice QCD is almost useless since inclusion of
chemical potential makes lattice action complex although there are several
new approaches have been proposed\cite{Nakamura}\cite{Glasgow}.
In this paper, we evaluate the transport coefficients by using statistical
ensembles generated by Ultra-Relativistic A-A collision simulator based on
Multiple Scattering Algorithm (URASiMA). Originally, URASiMA is an event
generator for the nuclear collision experiments based on the Multi-Chain
Model(MCM) of the hadrons\cite{Kumagai}.
Some of us(N.\ S and O.\ M) has already discussed thermodynamical properties
of a hot-dense hadronic state based on a molecular dynamical simulations of
URASiMA with periodic condition\cite{Sasaki1}. Recently, some groups have
been performed similar calculation with use of the different type of
event-generator UrQMD\cite{RQMD}, where Hagedorn-type temperature
saturation is reported. We improve URASiMA
to recover detailed balance at temperature below two hundred MeV.
As a result, Hagedorn-type behavior in
the temperature disappears\cite{Sasaki2}. This is the first calculation of
the transport coefficient of a hot and dense hadronic matter based on
an event generator.

In section 2, we review URASiMA and explain how to make ensembles with
finite
density
and finite temperature. Section 3 is devoted to the calculation for nucleon
diffusion constant through the first-kind fluctuation dissipation theorem.
Section 4 is concluding remarks.

\section{URASiMA for Statistical Ensembles} \label{URA}

URASiMA is a relativistic event generator based on hadronic multi-chain
model, which
aims at describing nuclear-nuclear collision by the superposition of
hadronic
collisions. Hadronic 2-body interactions are fundamental building blocks of
interactions in the model, and all parameters are so designed to reproduce
experimental data of hadron-hadron collisions.
Originally, URASiMA contains
2-body process (2 incident particle and 2 out-going particles), decay
process
(1 incident particle and 2 out-going particles), resonance (2 incident
particles and 1 out-going particle) and production process (2 incident
particles and n ($\ge$ 3) out going particles).
The production process is very important for the description of the multiple
production at high energies. On the other hand re-absorption
processes ( n ($ \ge $ 3) incident particles and 2 out-going particles)
thought to be unimportant in the collisions since system quickly expands
and they have not been included in the simulation.
On the other hand, in the generation of statistical ensembles in
equilibrium,
detailed balance between processes is essentially important. Lack of
re-absorption process leads one-way conversion of energy into particle
production rather than heat-up. As a result, artificial temperature
saturation occurs.

Therefore, role of re-absorption processes is very important
and we should take into account it.
However exact inclusion of multi-particle re-absorption processes is very
difficult.
In order to treat them effectively,
multi-particle productions and absorptions are treated as 2-body
processes including resonances with succeeding decays and/or preceding
formations of the resonances. Here two body decay and formation of
resonances are assumed. For example, $NN \to NN\pi$ is described as
$NN \to NR$ followed by decay of $R \to
N \pi$, where $R$ denotes resonance.
The reverse process of it is easily taken into account. In this approach,
all the known inelastic cross-sections for baryon-baryon interactions up to
$\sqrt{s} < 3$GeV, are reproduced.

\begin{center}
\begin{table}[b]
  \caption{
    Baryons, mesons and their resonances included in the URASiMA.
  }
  \label{tab:plst}
  \begin{tabular}{c|cccccccc}
\hline \hline
    nucleon  & $N_{938} $ & $N_{1440}$ & $N_{1520}$ & $N_{1535}$ &
               $N_{1650}$ & $N_{1675}$ & $N_{1680}$ & $N_{1720}$ \\
               \hline
    $\Delta$ & $\Delta_{1232}$ & $\Delta_{1600}$ &
               $\Delta_{1620}$ & $\Delta_{1700}$ &
               $\Delta_{1905}$ & $\Delta_{1910}$ &
               $\Delta_{1950}$ & \\ \hline
    meson & $\pi$ & $\eta$ & $\sigma_{800}$ & $\rho_{770}$ &&&& \\
\hline
  \end{tabular}
\end{table}
\end{center}

For the higher energy, $\sqrt{s} > 3$GeV, in order to give appropriate
total cross section, we need to take direct production process into account.
Only this point, detailed balance is broken in our simulation, nevertheless,
if temperature is much smaller than 3 GeV, the influence is negligibly
small.
For example , if the temperature of the system is 100 MeV , occurrence of
such process is suppressed by factor of $\mbox{exp}(-30)$ and thus time scale to
detect violation of detailed balance is very much longer than hadronic
scale.

In order to obtain equilibrium state, we put the system in a box and impose
periodic condition to URASiMA as the space-like boundary condition. Initial
distributions of particles are given by uniform random distribution of
baryons in a phase space. Total energy and
baryon number in the box are fixed at initial time and conserved through-out
simulation.  Though initial particles are only baryons, many masons are
produced through interactions. After thermalization time-period about
100 fm/c, system seems to be stationary.
In order to confirm the achievement of equilibrium, we calculate energy
distributions and particle numbers. Slope parameters of energy distribution
of all particles become the same value in the accuracy of statistics(Fig.\
1).
Thus, we may call this value as the temperature of the system. The fact that
numbers of species saturate indicates the achievement of chemical
equilibrium(Fig.\ 2). Running URASiMA many times with the same total energy
and total baryons in the box and taking the stationary configuration later
than $t=150$ fm/c, we obtain statistical ensemble with fixed
temperature and fixed baryon number(chemical potential).

\vspace{1cm}
\begin{center}
==============\\
fig.1 \\
==============\\

==============\\
fig.2 \\
==============\\
\end{center}
\noindent
By using the ensembles obtained through above mentioned manner, we can
evaluate thermodynamical quantities and equation of states\cite{Sasaki2}.

\section{Diffusion Constant} \label{Trans}

According to the Kubo's Linear Response Theory, the correlation of the
currents stands for admittance of the system(first fluctuation dissipation
theorem) and equivalently, random-force correlation gives impedance(Second
fluctuation dissipation theorem) \cite{Kubo}. As the simplest example, we
here focus our discussion to the diffusion constant. First
fluctuation dissipation theorem tells us that diffusion constant $D$ is
given by current(velocity) correlation,
\beq
D =\frac{1}{3} \int_{0}^{\infty}<\vev(t)\cdot \vev(t+t')> dt'. \label{eqn;fdt}
\eeq
Average $<\cdots>$ is given by,

\beq
<\cdots> = \frac{1}{\mbox{number of ensembles}}\sum_{\mbox{ensemble}}
\label{eqn;avg}
\frac{1}{\mbox{number of particle}}\sum_{\mbox{particle}} \cdots . \eeq
If the correlation decrease exponentially, i.e., \beq
<\vev(t)\cdot \vev(t+t')> \propto \exp{(- \frac{t'}{\tau})}, \label{eqn;rlx}
\eeq
with $\tau$ being relaxation time,
diffusion constant can be rewritten in the simple form, \beq
D = \frac{1}{3}<\vev(t)\cdot \vev(t)> \tau .\label{eqn;difcon} \eeq
Usually, diffusion equation is given as, \beq
{\frac{\partial}{\partial t}} f(t,\vex) = D \nabla^2 f(t,\vex),
\label{eqn;difeq}
\eeq
and diffusion constant $D$ has dimension of $[L^2 /T].$ Because of
relativistic nature of our system, we should use $\vebt = \frac{\vev}{c} =
\frac{\vep}{E}$ instead of $\vev$ in eq.(\ref{eqn;fdt}) and $D$ is obtained
by,

\beqa
D &=& \frac{1}{3}\int_{0}^{\infty}<\vebt(t)\cdot \vebt(t+t')> dt' c^2 .
\label{eqn;fdt2}
\\
&=&\frac{1}{3}<\vebt(t)\cdot \vebt(t)> c^2 \tau .\label{eqn;difcon2} \\
&=&\frac{1}{3}<\left(\frac{\vep(t)}{E(t)}\right)\cdot
\left(\frac{\vep(t)}{E(t)}\right)> c^2 \tau \eeqa
with $c$ being the velocity of light.
Figure 3 shows correlation function of the velocity of baryons. The figure
indicates that exponential damping is very good approximation.
Figure 4 displays the our results of baryon diffusion constant in a hot and
dense hadronic matter.
\begin{center}
==============\\
fig.3 \\
==============\\

==============\\
fig.4 \\
==============\\
\end{center}

Our results show clearer dependence on the baryon number density while
dependence on energy density is mild. This result means importance of
baryon-baryon collision process for the random walk of the baryons
and thus non-linear diffusion process of baryons occurs. In this sense, we
can state that
baryon number density in our system is still high. In the inhomogeneous
big-bang nucleosynthesis scenario, baryon-diffusion plays an important roll.
The leading part of the scenario is played by the difference
between proton diffusion and neutron diffusion\cite{inho}. In our
simulation,
strong interaction
dominates the system and we assume charge independence in the strong
interaction, hence, we can not discuss difference between proton and
neutron.
However obtained diffusion constant of baryon in our simulation can give
some
kind of restriction to the diffusion constants of both proton and neutron.

From diffusion constant, we can calculate charge
conductivity\cite{textbook}.
Figure 5 shows baryon number conductivity $\sigma_{\rm B}$, \beq
\sigma_{\rm B} = \displaystyle{\frac{n_{\rm B}}{k_{\rm B}T}}D, \eeq
where $n_{\rm B}$ is baryon number density, $T$ is temperature and $k_{\rm
B}$ is Boltzmann constant(put as unity through out this paper),
respectively.

\begin{center}
==============\\
fig.5 \\
==============\\
\end{center}
\noindent
Therefore, if we want, we can discuss Joule heat and entropy production in
the
{\it Baryonic circuit} based on the above baryon number conductivity.

Because fundamental system in URASiMA is high energy
hadronic collisions, we use relativistic notations usually. However,
diffusion equation
(\ref{eqn;difeq}) is not Lorentz covariant and is available only on the
special system i.e. local rest frame of the thermal medium. For the
full-relativistic description of the space-time evolution of a hot and
dense matter, we need to establish relativistic Navier-Stokes
equation\cite{Namiki}.
Taking correlation of appropriate currents, we can easily evaluate
viscosities and heat conductivity in the same manner \cite{Sasaki3} .

\section{Concluding Remarks} \label{conc}

Making use of statistical ensembles obtained by an event generator URASiMA,
we evaluate diffusion constants of baryons in the hot and dense hadronic
matter. Our results show strong dependence on baryon number density and
weak dependence on temperature.
The temperature in our simulation is limited only small range, i.e., from
100 MeV to 200 MeV, and this fact can be one of the reasons  why the change
of
diffusion constant of temperature is not clear.
Strong baryon number density dependence indicates that, for the baryon
diffusion process, baryon plays more important roll than light mesons. In this
sense our simulation corresponds to high density region and non-linear
diffusion process occurs. Calculation of the diffusion constants is the
simplest examples of first fluctuation dissipation theorem.
In principle, taking correlation of appropriate currents, i.e. energy
flow, baryon number current, stress-tensor, etc.,
we can evaluate any kinds of transport
coefficients.
However, in relativistic transport theory, there exist several delicate
points, e.g., relativistic property makes difference of mass and energy
meaningless and, as a result, meaning of the "flow" of the fluid and "heat
flow" become ambiguous\cite{Namiki}\cite{Landau}.
The choice of the current depends on the macroscopic phenomenological
equations which contain transport coefficients. Once we establish
phenomenological equations for the high temperature and high density
hadronic
matter, we can evaluate the appropriate transport coefficients in the same
manner. Detailed discussion will be reported in our forthcoming
paper\cite{Sasaki3}.

\noindent
{\bf Acknowledgment}

The authors would like to thank prof. T.\ Kunihiro for the discussion. 
This work is supported by Grant-in-Aid for scientific research number
11440080 by Monbusho.  Calculation has been done at Institute for Nonlinear
Sciences and Applied Mathematics, Hiroshima University.


\newpage
\begin{figure}
 \includegraphics[scale=0.80]{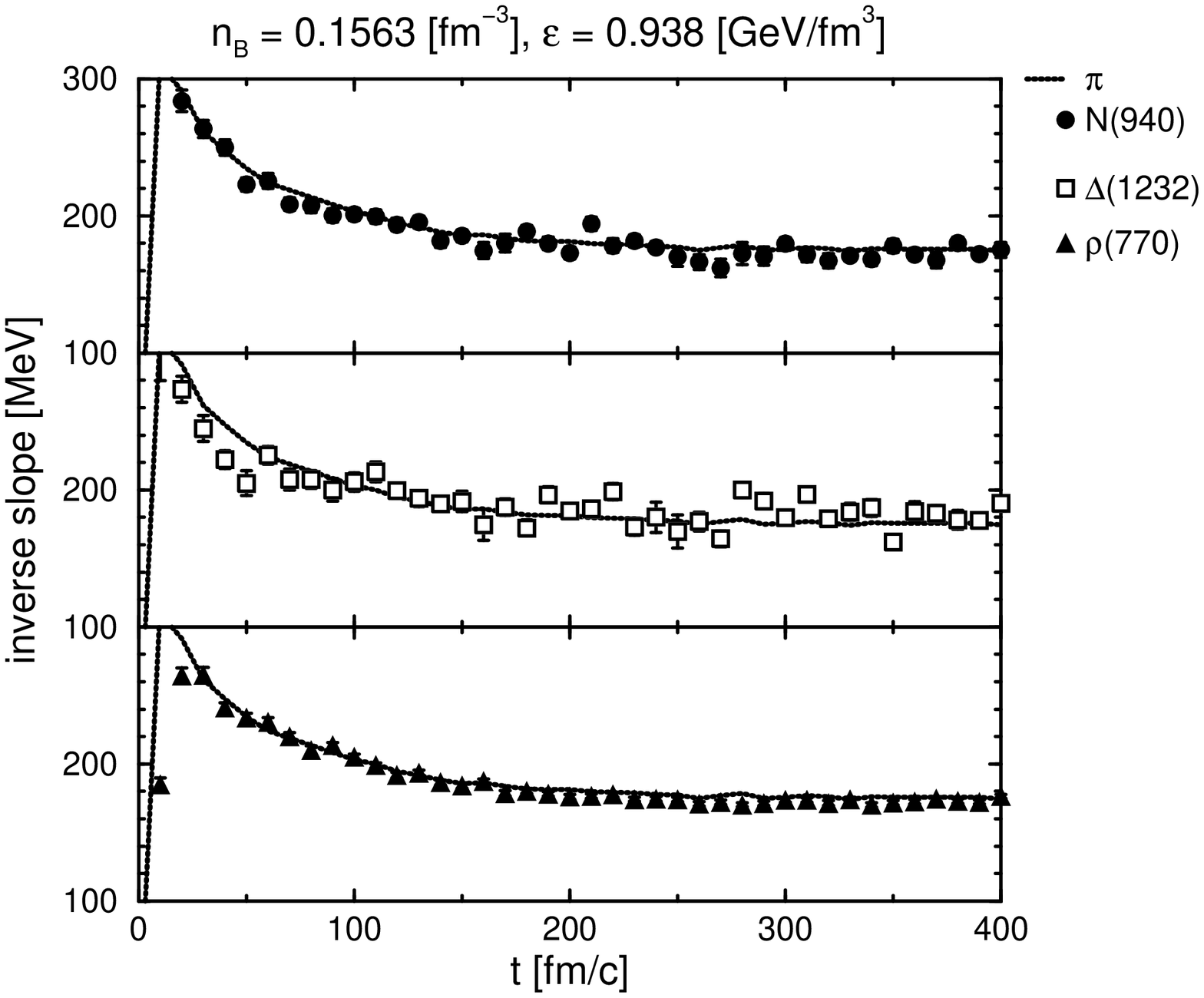}

\caption[slope]{
The time evolution of the inverse slopes $\beta^{-1}$ of $N_{938}$,
$\Delta_{1232}$, $\rho_{770}$ and $\pi$ at $n_{B}=0.1563\;\mbox{fm}^{-3}$ 
and $\varepsilon_{tot}=0.938\;\mbox{GeV/fm}^{3}$.
$\beta^{-1}$ is obtained by the energy distributions,
$
      \frac{dN}{d^{3}\vec{p}} = \frac{dN}{4\pi Ep dE}
      = C\exp(-\beta E).
$
The dotted line stands for the $\beta^{-1}$ of $\pi$.
  }

\label{fig:slope}
\end{figure}

\newpage
\begin{figure}
  \includegraphics[scale=0.80]{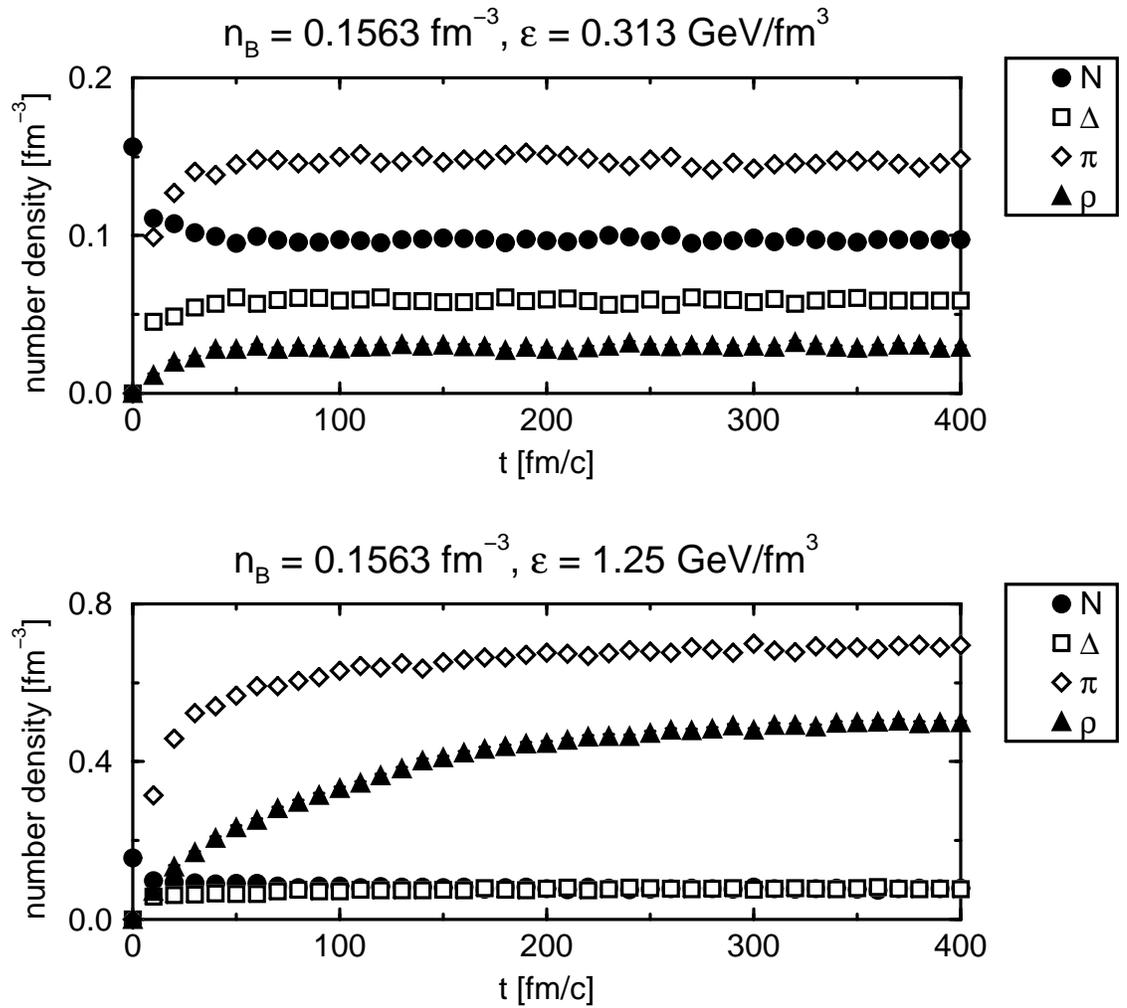}
  \caption{
    The time evolution of number densities: (a) $\varepsilon_{tot}=0.313\;
    [\mbox{GeV/fm}^{3}]$ and (b) $\varepsilon_{tot}=1.25\;
    [\mbox{GeV/fm}^{3}]$.
}
\label{fig:pdens}
\end{figure}

\newpage
\begin{figure}
  \includegraphics[scale=0.8]{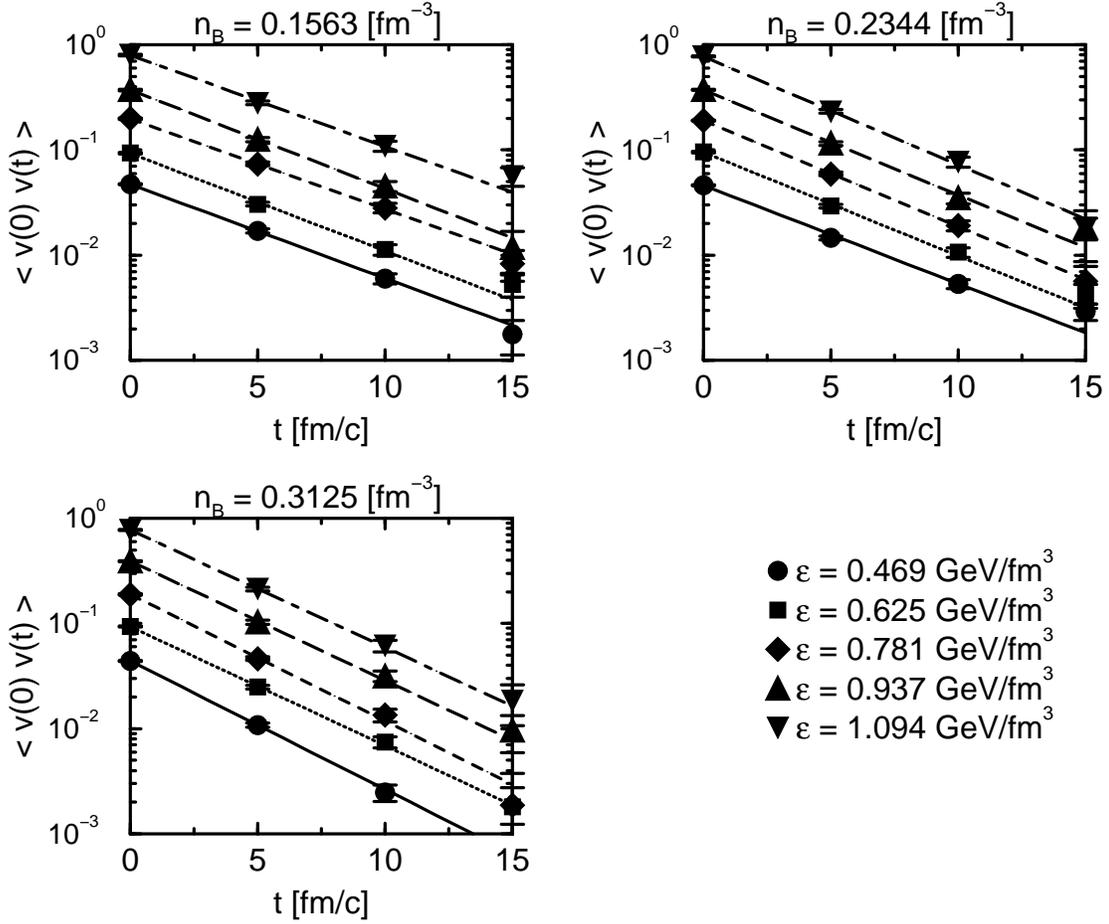}
  \caption{
    Velocity correlation of the baryons as a function of time.
    Lines correspond to the fitted results by exponential function.
    Normalizations of the data are arbitrary.
}
\end{figure}

\newpage
\begin{figure}
  \includegraphics[scale=0.80]{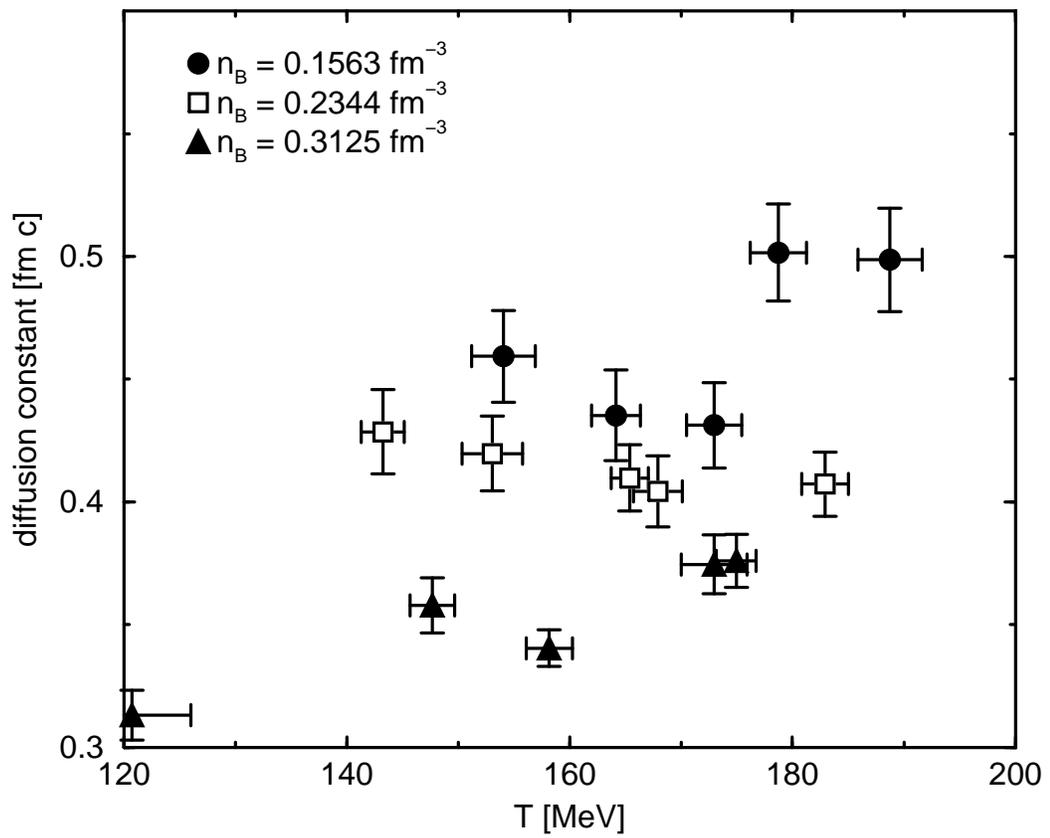}
  \caption{ Diffusion constant of baryons.
   }
\label{fig:diff}
\end{figure}

\newpage
\begin{figure}
  \includegraphics[scale=0.80]{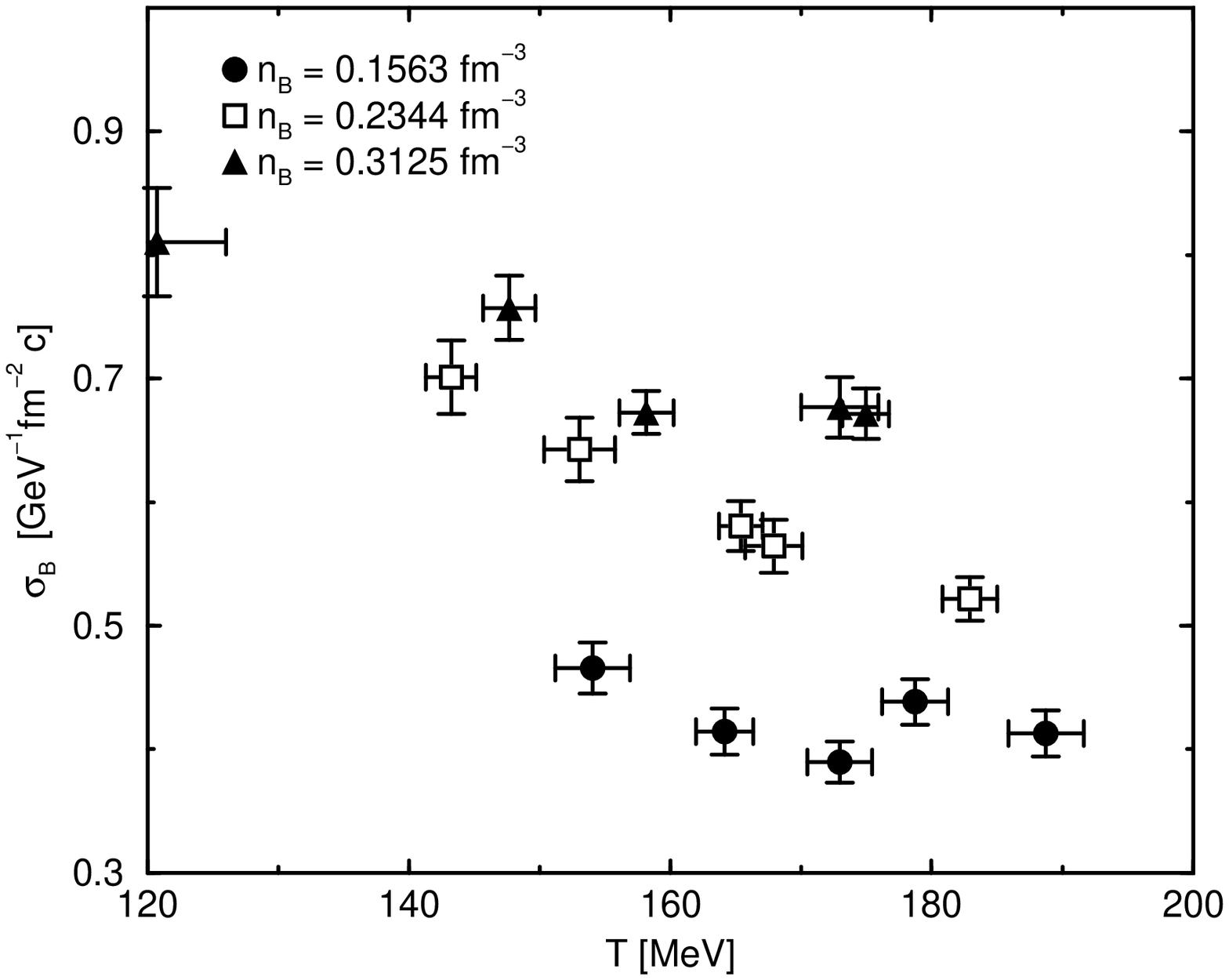}
  \caption{Baryon charge conductivity.
}
\label{fig:cond}
\end{figure}


\baselineskip 12pt
\begin{thebibliography}{99}
\bibitem{QM99}For example, see the proceedings of Quark Matter '97, Nucl.\
Phys.,{\bf A638}(1998)1c.
\bibitem{Sakai} S.\ Sakai, A.\ Nakamura and T.\ Saito, Nucl.\ Phys.,{\bf
A638}(1997)535c.
\bibitem{Nakamura} A.\ Nakamura, Acta Phys.\ Pol.,{\bf B16}(1985)635; P.\
Harsenfratz, F.\ Kersh and I.\ O.\ Stamatesucu, Phys.\ Lett.\
{\bf 133B}(1983)221.
\bibitem{Glasgow}
I.\ M.\ Barbour for UKQCD Collaboration, Nucl.\ Phys.\ {\bf A642}(1998)251;
J.\ Engels, O.\ Kaczmarek, F.\ Karsch, E.\ Laermann, in the proceedings of 
17th International Symposium on Lattice Field Theory (LATTICE 99),
hep-lat/9908046.
\bibitem{Kumagai} S.\ Dat{\'e}, K.\ Kumagai, O.\ Miyamura and X.\ Z.\ Zhang,
JPSJ {\bf 64}(1995)766.
\bibitem{Sasaki1} N.\ Sasaki and O.\ Miyamura, Prog.\ Theor.\ Phys.\ Suppl.
{\bf 129}(1997)39.
\bibitem{RQMD}S.\ A.\ Bass et al., Prog.\ Part.\ Nucl.\ Phys.\ {\bf
41}(1998)225; M.\ Belkacem et al.,Phys.\ Rev.\ {\bf C58}(1998)1727.
\bibitem{Sasaki2}
N.\ Sasaki, in preparation.
\bibitem{Kubo}R.Kubo, Reports on Progress in Physics
{\bf 29}Part I (1966)255.
\bibitem{inho}I.\-S.\ Suh and G.\ J.\ Mathews, Phys.\ Rev.\ {\bf D58}(1998)
3002.
\bibitem{textbook}M.\ Toda, R.\ Kubo and N.\ Saito, {\it Statistical Physics I}
, (Springer-Verlag, Berlin, 1992); R.\ Kubo, M.\ Toda and N.\ Hashitsume,
{\it Statistical Physics II}, (Springer-Verlag, Berlin, 1991).
\bibitem{Namiki} M.\ Namiki and C.\ Iso, Prog.\ Theor.\ Phys.\
{\bf 18}(1957)591; C.\ Iso, K.\ Mori and M.\ Namiki, Prog.\ Theor.\ Phys.\
{\bf 22}(1959)403.
\bibitem{Sasaki3}N.\ Sasaki, O.\ Miyamura, S.\ Muroya and C.\ Nonaka,
in preparation.
\bibitem{Landau} L.\ D.\ Landau and E.\ M.\ Lifsitz, {\it Fluid Mechanics,
Pergamon  Press., Oxford(1989)}; S. Weinberg, Astrophys.\ J.\ {\bf
168}(1971)175.

\end{thebibliography}
\end{document}